\documentclass[prl,twocolumn,showpacs,floatfix,amsmath,amssymb,superscriptaddress]{revtex4-2}
\usepackage{amsfonts}
\usepackage{stmaryrd}
\usepackage{bbm}
\usepackage{mathrsfs}
\usepackage{tipa}
\usepackage{amssymb}
\usepackage{txfonts}
\usepackage{graphicx}
\usepackage{dcolumn}
\usepackage{epstopdf}
\usepackage[colorlinks,linkcolor=black,urlcolor=black,citecolor=blue]{hyperref}
\usepackage{multirow}
\usepackage{subfigure}
\usepackage{url}
\usepackage{upgreek}
\usepackage[utf8]{inputenc}
\usepackage[english]{babel}
\usepackage[locale = US]{siunitx}
\usepackage{lipsum}
\usepackage{comment}

\def\braket#1{\langle #1 \rangle}

\begin{document}

\title{Ellipticity control of terahertz high-harmonic generation in a Dirac semimetal}

\author{Semyon~Germanskiy}
\email{germanskiy@ph2.uni-koeln.de}
\affiliation{Institute of Physics II, University of Cologne, 50937 Cologne, Germany}

\author{Renato~M.~A.~Dantas}
\email{renatomiguel.alvesdantas@unibas.ch}
\affiliation{Max Planck Institute for the Physics of Complex Systems, 01187 Dresden, Germany}
\affiliation{Department of Physics, University of Basel, 4056 Basel, Switzerland}

\author{Sergey~Kovalev}
\affiliation{Institute of Radiation Physics, Helmholtz-Zentrum Dresden-Rossendorf, 01328 Dresden, Germany}

\author{Chris~Reinhoffer}
\author{Evgeny~A.~Mashkovich}
\author{Paul~H.~M.~van~Loosdrecht}
\affiliation{Institute of Physics II, University of Cologne, 50937 Cologne, Germany}

\author{Yunkun~Yang}
\author{Faxian~Xiu}
\affiliation{State Key Laboratory of Surface Physics and Department of Physics, Fudan University, Shanghai, China}
\affiliation{Shanghai Qi Zhi Institute, No. 701 Yunjin Road, Xuhui District, Shanghai, 200232, China}

\author{Piotr~Sur\'{o}wka}
\affiliation{Department of Theoretical Physics, Wroc\l{}aw University of Science and Technology, 50-370 Wroc\l{}aw, Poland}
\affiliation{Institute for Theoretical Physics, University of Amsterdam, 1090 GL Amsterdam, The Netherlands}
\affiliation{Dutch Institute for Emergent Phenomena (DIEP), University of Amsterdam, 1090 GL Amsterdam, The Netherlands}
\affiliation{Max Planck Institute for the Physics of Complex Systems, 01187 Dresden, Germany}

\author{Roderich~Moessner}
\affiliation{Max Planck Institute for the Physics of Complex Systems, 01187 Dresden, Germany}

\author{Takashi~Oka}
\affiliation{The Institute for Solid State Physics, The University of Tokyo, Kashiwa, Chiba 277-8581, Japan}
\affiliation{Max Planck Institute for the Physics of Complex Systems, 01187 Dresden, Germany}
\affiliation{Max Planck Institute for Chemical Physics of Solids, 01187 Dresden, Germany}

\author{Zhe~Wang}
\email{zhe.wang@tu-dortmund.de}
\affiliation{Department of Physics, TU Dortmund University, 44227 Dortmund, Germany}
\affiliation{Institute of Physics II, University of Cologne, 50937 Cologne, Germany}

\date{\today}

\begin{abstract}
We report on terahertz high-harmonic generation in a Dirac semimetal as a function of the driving-pulse ellipticity and on a theoretical study of the field-driven intraband kinetics of massless Dirac fermions.
Very efficient control of third-harmonic yield and polarization state is achieved in electron-doped Cd$_3$As$_2$ thin films at room temperature. The observed tunability is understood as resulting from terahertz-field driven intraband kinetics of the Dirac fermions.
Our study paves the way for exploiting nonlinear optical properties of Dirac matter for applications in signal processing and optical communications.
\end{abstract}

\maketitle

Strong-field driven nonlinear response provides a fruitful path for the discovery and understanding of intriguing dynamical processes of quantum matter \cite{CorkumKrausz,GhimireReis19,Chang20}.
The dependence of high-order harmonic generation on the ellipticity of the driving laser exhibits characteristic features of the nonlinear dynamics
\cite{Sansone06,Ghimire2011,Vampa2015,Cireasa15,Kfir15,
Fan15,Liu2017,You2017,Yoshikawa17,Taucer17,
TancogneDejean2017,Luo19,Rubio2021}.
The polarization state of electromagnetic waves has not only been employed extensively to investigate fundamental physical properties of matter, which for linear response underlies numerous spectroscopic techniques, but also been demonstrated to be very efficient in controlling nonequilibrium states of matter and their nonlinear response via strong light-matter interactions.
A plethora of very interesting nonlinear physical phenomena have been found in different states of matter, i.e. gases \cite{CorkumKrausz,Chang20}, liquids \cite{Zhang21}, and solids \cite{GhimireReis19,Chang20}. In particular, high-order harmonic generation (HHG) is found to exhibit characteristics of nonlinear response in atomic or molecular gases \cite{Ferray88,Rhodes88,Budil1993,Corkum94} as well as in solid-state materials (see e.g. \cite{Ghimire2011,Schubert2014,Vampa2015,Liu2017,You2017,
Yoshikawa17,Taucer17,Hafez2018,Soavi2018,Cheng20,
Kovalev2020,Chu2020,Kovalev21,Lim2021,Mao2022}).

Decades ago high-harmonic generation was observed when driving noble gases with picosecond laser radiation \cite{Rhodes88,Budil1993,Corkum94}. The coherent radiation emitted in strong-field driven atomic and molecular gases has enabled spectroscopic studies in the extreme ultraviolet and soft X-ray regimes and also for ultrafast dynamics on attosecond time scales \cite{CorkumKrausz,Chang20}.
The yield of high harmonics in atomic gases is maximized for linearly polarized lasers, but drops rapidly with increasing ellipticity of the driving laser, already by two orders of magnitude at a relatively small ellipticity $\epsilon_f \lesssim 0.5$ (see e.g. \cite{Budil1993,Corkum94}).
The strong ellipticity dependence of HHG provides tremendous opportunities for applications, including production of isolated attosecond pulses \cite{Sansone06}, probing chiral interactions of molecules through sub-femtosecond electronic dynamics \cite{Cireasa15}, and detection of soft X-ray magnetic circular dichroism in magnetic substances \cite{Kfir15,Fan15}.

In contrast to the atomic gases where the HHG is well understood in terms of three-step processes (ionization, acceleration, and recollision) \cite{Lewenstein94,CorkumKrausz}, solid-state materials exhibit versatile ellipticity dependence of HHG \cite{Ghimire2011,Vampa2015,Liu2017,You2017,
Yoshikawa17,Taucer17}.
On the one hand, atomic-like ellipticity dependence was also found in some solids (e.g. rare-gas solids \cite{Vampa2015}, ZnO crystal \cite{Ghimire2011}, monolayer MoS$_2$ \cite{Liu2017}). On the other hand, enhanced harmonic yield can be realized at larger ellipticity reaching a maximum for circularly polarization ($\epsilon_f = 1$) such as in MgO \cite{You2017}, or at finite ellipticity, e.g. $\epsilon_f=0.32$ in graphene \cite{Yoshikawa17}.
The ellipticity dependence appears to also vary with driving-pulse energy, suggesting frequency-dependent mechanisms. For example, graphene exhibits the unusual ellipticity dependence for a driving-pulse energy of 0.26~eV \cite{Yoshikawa17}, whereas for slightly higher energies 0.32 and 0.4~eV an atomic-like behavior is restored \cite{Taucer17}. 

\begin{figure*}[t]
\includegraphics[width=500pt]{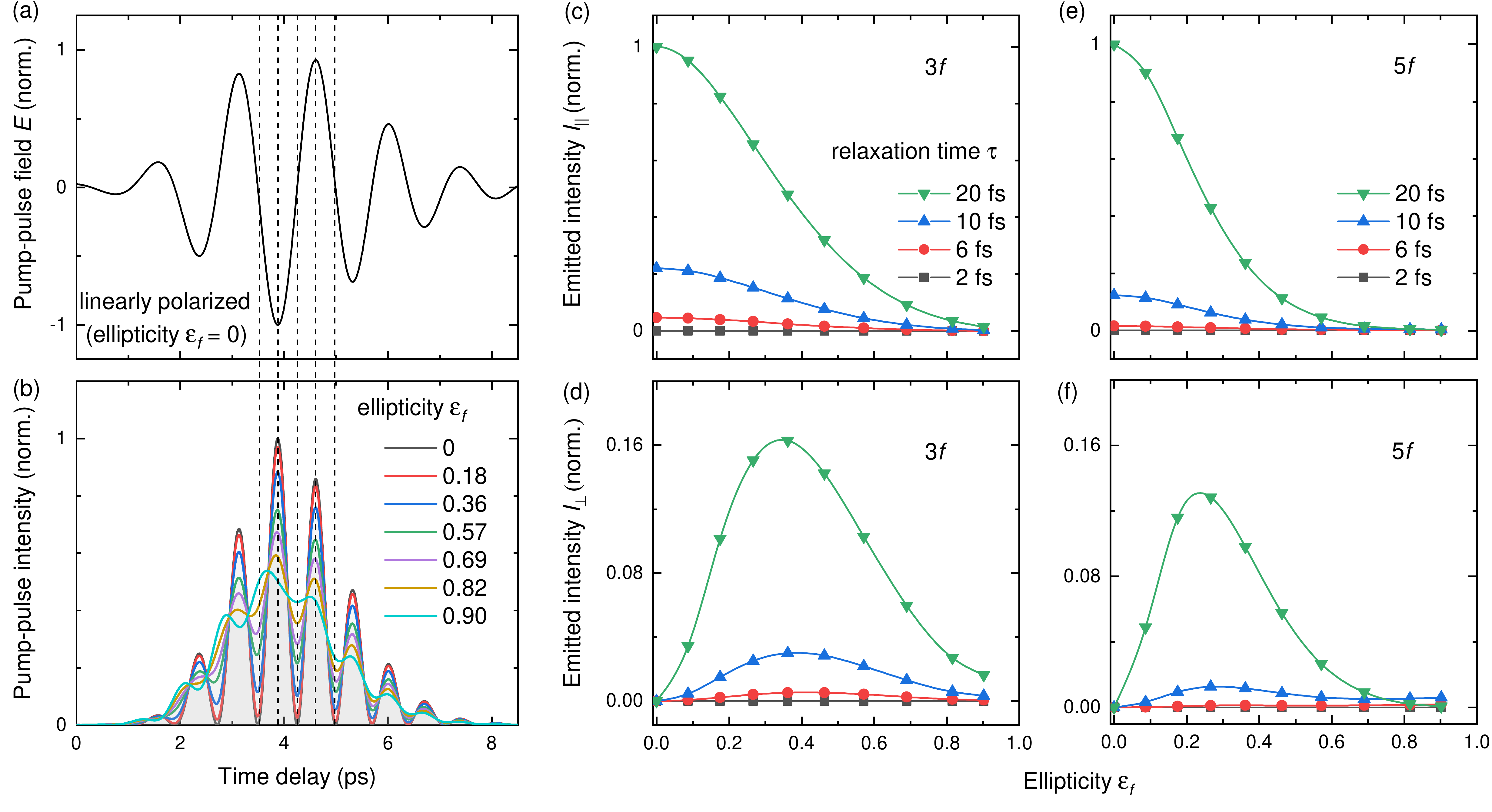}
\caption{\label{fig:theoryHHG}
(a) An idealized driving pulse with linear polarization, a central frequency of $f=0.69$~THz, and full width at half maximum (FWHM) of 0.11~THz. 
(b) Intensity of driving pulse for various ellipticities $\varepsilon_f$.
Intensity of emitted third-order harmonic radiation ($3f$), for (c) parallel $I_\parallel$ and (d) perpendicular $I_\perp$ components, as a function of the ellipticity for various relaxation times $\tau=2$, 6, 10, and 20~fs.
The intensity is normalized to the maximum value of $I_\parallel$.
Intensity of emitted fifth-order harmonic radiation ($5f$) versus ellipticity for (e) parallel and (f) perpendicular components.
}
\end{figure*}

A crucial role of interband excitations in determining the ellipticity dependence of HHG has been emphasized by previous studies (see e.g. \cite{TancogneDejean2017,Yoshikawa17,Luo19,Rubio2021}).
The experimentally observed ellipticity dependencies can result from combined effects of interband excitations with dynamical Bloch oscillations \cite{Luo19}, nonlinear coupling to intraband excitations \cite{TancogneDejean2017,Rubio2021}, or other quantum mechanical effects (e.g. Zener tunneling) \cite{Yoshikawa17}.
In this work, without the need for the complex interband processes, we investigate ellipticity dependence of HHG in a very different but very representative setting, in which the field-driven intraband kinetics of massless Dirac fermions is primarily responsible for the ellipticity dependence.
We study HHG of terahertz (THz) field-driven relativistic quasiparticles in an electron-doped Dirac semimetal, where the interband transitions are essentially Pauli-blocked due to the low energy of THz photons. 
By measuring THz third-harmonic generation (THG) in the well-established three-dimensional Dirac semimetal Cd$_3$As$_2$, we find an evident dependence of the THG ellipticity and intensity on the driving-pulse ellipticity, in good agreement with our results based on kinetic theory. Our work provides a very efficient approach to control the THz HHG and reveals the underlying nonlinear kinetics. 

\begin{figure*}
\includegraphics[width=500pt]{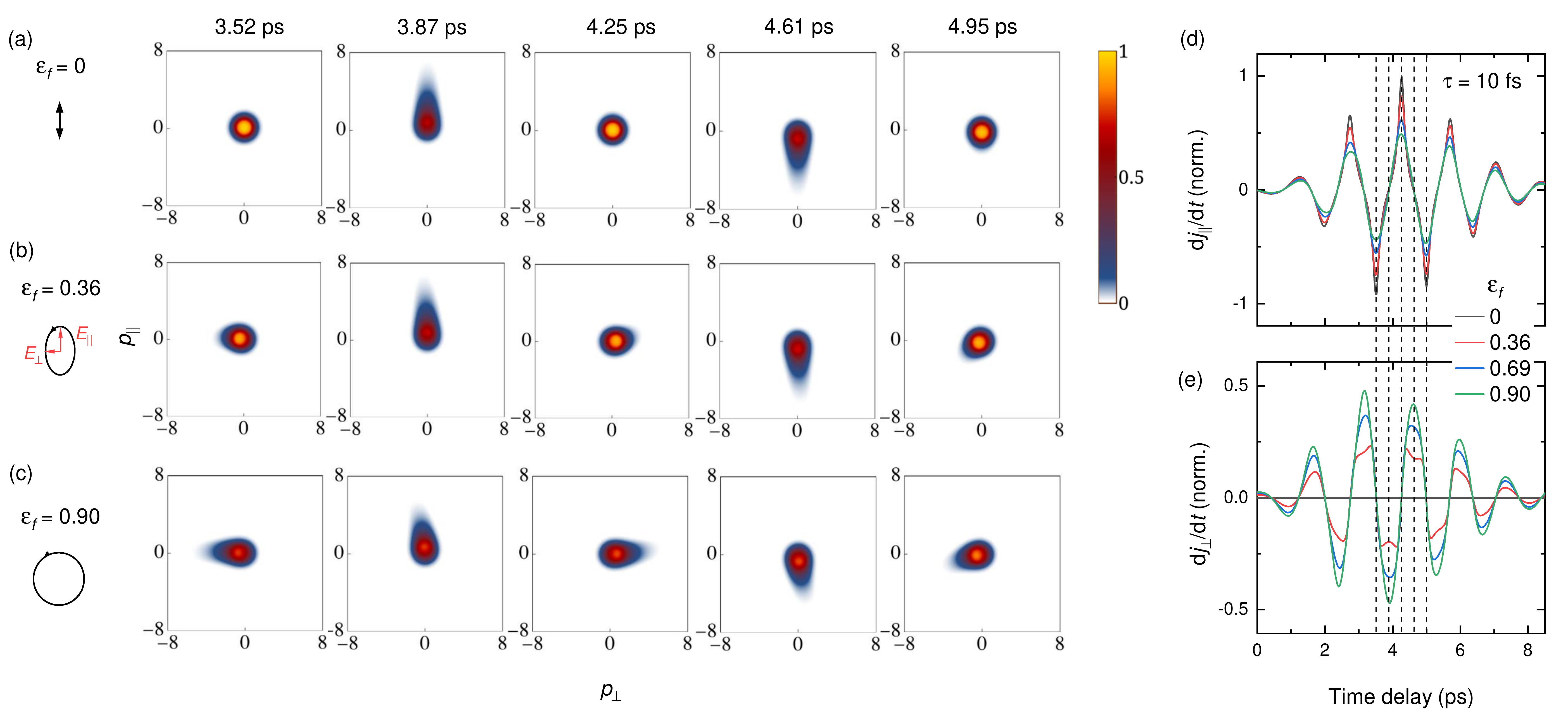}
\caption{\label{fig:Distribution}
Projection of the electron distribution function on the ($p_\perp$,$p_\parallel$) momentum plane at selected time delays 3.52, 3.87, 4.25, 4.51, and 4.95 ps [corresponding to the dashed lines in (d)(e) and in Fig.~\ref{fig:theoryHHG}(a)(b)] for a relaxation time $\tau=10$~fs and various ellipticities (a) $\varepsilon_f=0$, (b) 0.36, and (c) 0.90.
The momenta are given in units of the Fermi momentum $E_F/v_F$.
The time-derivative of the current density (d) $dj_\parallel/dt$ and (e) $dj_\perp/dt$ as a function of time delay, which is normalized to the maximum value of $dj_\parallel/dt$.
}
\end{figure*}

We start with a theoretical analysis of the THz driven kinetics in a single Dirac-electron band by Boltzmann transport theory.
While the initial equilibrium state is defined by the Fermi-Dirac distribution $f_0(\mathbf{p})$ at room temperature, we evaluate the time evolution of the distribution function $f(t,\mathbf{p})$ under the drive of an external THz field $\mathbf{E}(t)$ via the Boltzmann equation
\begin{equation}
\left(\frac{\partial}{\partial t}+\frac{1}{\tau}\right)f(t,\mathbf{p})-e\mathbf{E}(t)\cdot\nabla_\mathbf{p}f(t,\mathbf{p})=\frac{1}{\tau}f_0(\mathbf{p}),
\end{equation}
where $\tau$ is a characteristic relaxation time for intraband processes and $e$ denotes the elementary charge. 
Idealized THz pulses with linear or elliptical polarizations [see Fig.~\ref{fig:theoryHHG}(a),(b)] are adopted to simulate the driven intraband processes of the relativistic quasiparticles in the time domain.
Ellipticity $\varepsilon_f$ of the terahertz pulses was computed via
$\varepsilon = \tan\Bigl[ \frac{ 1 }{2}\arcsin(S_3/S_0) \Bigr]$
with the Stokes parameters given by
$S_0 = \left\langle\vert\hat{E}_\perp\hat{E}_\perp^*\vert + \vert\hat{E}_\parallel\hat{E}_\parallel^*\vert \right\rangle$ and
    $S_3 = \left\langle 2\text{Im}(\hat{E}^*_\perp\hat{E}_\parallel)\right\rangle$, where $\hat{E}_\parallel(t)$ and $\hat{E}_\perp(t)$ are the THz electric-field components and \(\left\langle \dots\right\rangle\) represents time averaging \cite{Gabor1946, MaxBorn2019}.
The ellipticity can be roughly interpreted as the ratio between the maximum electric-field components, $E_\perp$ and $E_\parallel$, along the minor- and major-axes of the polarization ellipse, respectively [see Fig.\ref{fig:Distribution}(b) for an illustration].
In the following, perpendicular or parallel components of other quantities are defined in a similar way,
as perpendicular or parallel to the linear polarization ($\varepsilon_f=0$), respectively.

\begin{figure*}
\includegraphics[width=500pt]{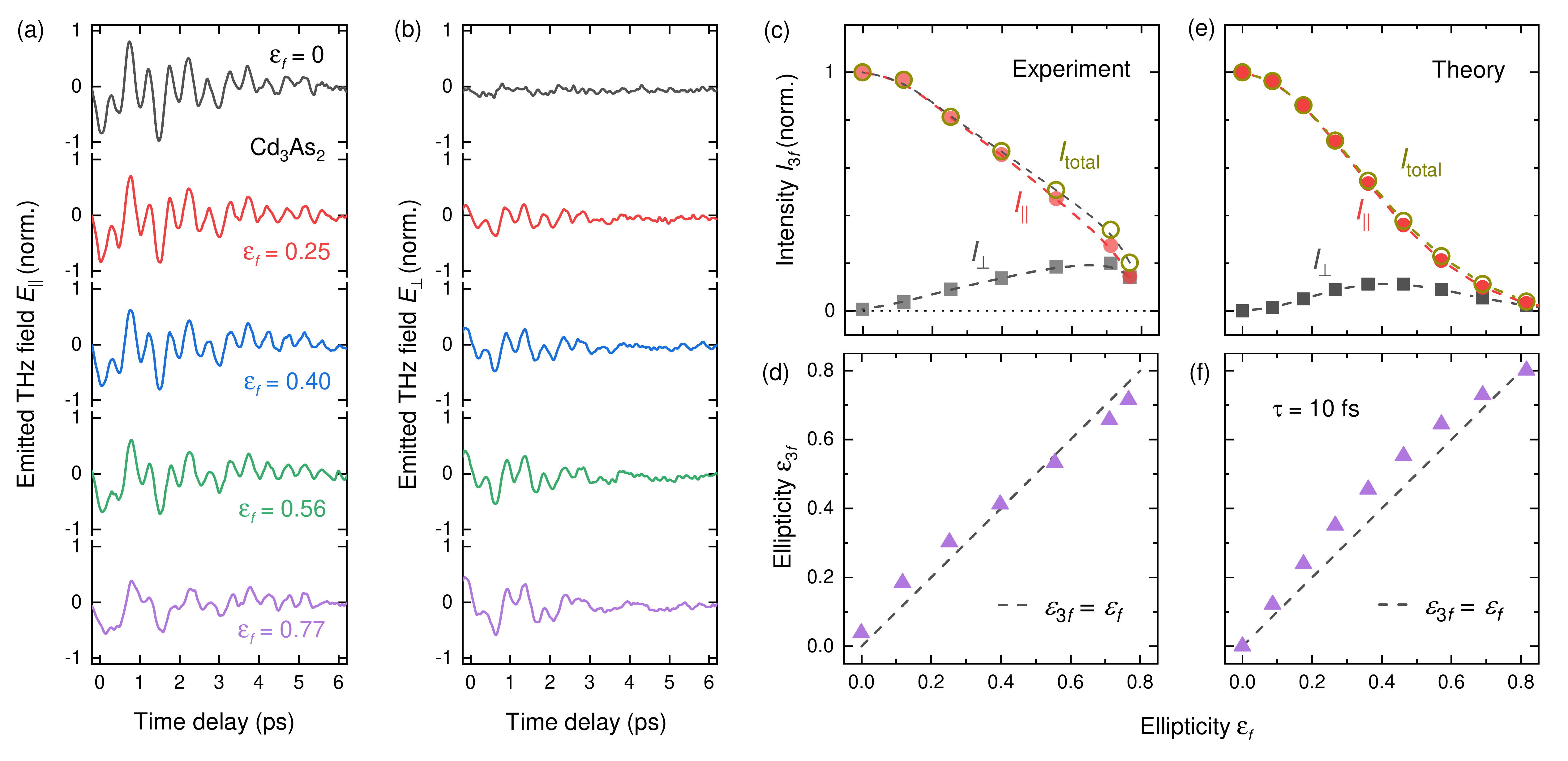}
\caption{\label{fig:ellipDepExp}
Electric-field components (a) $E_\parallel$ and (b) $E_\perp$ of emitted terahertz radiation from Cd$_3$As$_2$ at room temperature.
The data are normalized to the maximum value of $E_\parallel$ at $\varepsilon_f=0$.
(c) Intensity $I_{3f}$ of terahertz THG from Cd$_3$As$_2$ as a function of ellipticity at room temperature. The dashed lines are guides to the eyes.
(d) Ellipticity $\varepsilon_{3f}$ of observed THG  versus driving-pulse ellipticity $\varepsilon_f$. The dashed line depicts $\varepsilon_{3f}=\varepsilon_f$.
(e) Intensity and (f) ellipticity of theoretically obtained THG versus driving-pulse ellipticity $\varepsilon_f$ for $\tau=10$~fs.
}
\end{figure*}

The analytical solution for the Boltzmann equation [Eq.(1)] satisfying the boundary condition $f(0,\mathbf{p})=f_0(\mathbf{p})$ is given by~\cite{Dantas21,pnas.2200367119}
%------------------------------------------------------
\begin{align}
f \left( t, \mathbf{p}\right)= {}& \exp(-\tfrac{t}{\tau}) f_0 \left( \mathbf{p} - e \mathbf{\Delta} (t,0) \right) \nonumber \\
{}&+\frac{1}{\tau}  \int^{t}_0 ds\, \exp(\tfrac{s-t}{\tau}) f_0 \left( \mathbf{p} - e \mathbf{\Delta} (t,s)\right),
\label{eq:ASBE}
\end{align}
%------------------------------------------------------
where $\mathbf{\Delta} (t,s)=-\int^t_s  \mathbf{E} (\tilde{s}) \, d\tilde{s}$. 
The current density is defined as
%------------------------------------------------------
\begin{equation}
\mathbf{j}(t)= - 2 \, e \bigg[ \exp{\left(-\tfrac{t}{\tau}\right)}  \braket{ \mathbf{v} (t,0) }  + \int^{t}_0 \tfrac{ds}{\tau} \exp{\left(\tfrac{s-t}{ \tau}\right)}  \braket{\mathbf{v} (t,s)} \bigg], 
\end{equation}
%------------------------------------------------------
where $\braket{\mathbf{v}(t,s)} = \int\frac{d^3p}{(2\pi\hbar)^3} \mathbf{v}_{\mathbf{p}}  f_0(\mathbf{p} -e\mathbf{\Delta} (t,s))$ corresponds to the expectation value of the group velocity of the Dirac fermions.
The electric field of the emitted THz radiation is proportional to the time derivative of the current density. Through a Fourier transformation of the time-domain data, we compute the HHG intensity.
The obtained intensity of the emitted third-harmonic $3f$ and fifth-harmonic $5f$ radiation versus ellipticity is shown in Fig.~\ref{fig:theoryHHG}(c)-(f) for various relaxation times $\tau$, a peak electric field of 212~kV/cm and typical values of Fermi energy $E_F=118$~meV and Fermi velocity $v_F=7.8\times10^5$~m/s in a Dirac semimetal, Cd$_3$As$_2$ \cite{Kovalev2020}.

The obtained parallel $I_\parallel$ and perpendicular $I_\perp$ intensity components exhibit clearly different ellipticity dependence for every relaxation time and harmonic. Whereas $I_\parallel$ drops monotonically with increasing ellipticity, an initial increase of $I_\perp$ is followed by a continuous decrease approaching the circular polarization, exhibiting a maximum at a finite ellipticity. For $\tau=20$~fs, the maximum of the THG occurs at $\varepsilon_f^{\text{max}}=0.36$, while at $\varepsilon_f^{\text{max}}=0.24$ for the fifth-harmonic generation (FHG).
For both harmonics the position of the maximum shifts towards larger ellipticity with decreasing relaxation time.
Moreover, for different harmonics or relaxation times, $I_\perp$ is always considerably smaller than  $I_\parallel$, thus the total intensity decreases continuously with increasing ellipticity.
This is clearly in contrast to an enhanced THG at finite ellipticity due to interband excitations involved \cite{Yoshikawa17}. 

To experimentally study the nonlinear response due to intraband processes, we measure THz-driven third harmonic generation from electronically doped Cd$_3$As$_2$ thin films, a well-established three-dimensional Dirac semimetal \cite{Zhijun13,Borisenko14,Neupane2014,Liu2014}.
High-quality thin films of Cd$_3$As$_2$ with a typical thinkness of 120~nm were grown by molecular beam epitaxy, as described in Ref.~\cite{Liu2015} in detail.

Intense THz radiation is generated based on optical rectification of laser pulses (80~fs, 4~mJ, 800~nm) in
a LiNbO$_3$ crystal using a conventional tilted-pulse-front scheme (see e.g. \cite{Hebling2002,Hirori2011}).
Narrow-band multicycle THz driving pulses with a peak field of 130~kV/cm, a central frequency around $f=0.67$~THz and linewidth of 0.15~THz were obtained through a band-pass filter.
An \textit{x}-cut single crystalline quartz with a thickness of 2.18~mm was adopted to tune the polarization state of terahertz driving pulses.
The experimentally obtained driving pulses for various ellipticities are displayed in Fig. S1(a)\cite{suppl2022}, with a largest achieved ellipticity of $\varepsilon_f=0.77$.
Emitted THz pulses from a Cd$_3$As$_2$ thin film in a transmission configuration were detected via electro-optic sampling in a $\langle 110 \rangle$-cut GaP crystal.
The electric field of the emitted radiation from a Cd$_3$As$_2$ thin film at room temperature was recorded through a $3f$ band-pass filter as a function of time delay.
The parallel $E_\parallel(t)$ and perpendicular $E_\perp(t)$ components are measured separately by using THz wire-grid polarizers, which are presented in Fig.~\ref{fig:ellipDepExp}(a) and Fig.~\ref{fig:ellipDepExp}(b), respectively, for various ellipticities from $\varepsilon_f=0$ to $0.77$.

For a linearly polarized driving pulse, $E_\parallel(t)$ exhibits strong oscillations corresponding to the frequency of $3f$ (see Fig.~S1(b)~\cite{suppl2022} for the spectrum in frequency domain), whereas $E_\perp(t)$ is almost zero.
At finite ellipticity $E_\perp(t)$ starts to increase, and exhibits also the $3f$ oscillations. In contrast, the parallel component $E_\parallel(t)$ decreases continuously with enhanced ellipticity.
The obtained THG intensity and ellipticity is shown in Fig.~\ref{fig:ellipDepExp}(c)(d) as a function of ellipticity.
For comparison theoretical results for the same peak field and $\tau=10$~fs are shown in Fig.~\ref{fig:ellipDepExp}(e)(f).

Experimentally we observe that $I_\parallel$ decreases continuously with increasing ellipticity, whereas $I_\perp$ exhibits a broad maximum around $\varepsilon_f^{\text{max}}=0.6$.
These ellipticity dependence qualitatively agrees very well with the theory results [Fig.~\ref{fig:ellipDepExp}(e)], apart from the quantitative difference on $\varepsilon_f^{\text{max}}$.
Moreover, the observed $I_\perp$ is notably smaller than $I_\parallel$, leading to a monotonic drop of the total THG intensity $I_{\text{total}}$ with increased ellipticity [Fig.~\ref{fig:ellipDepExp}(c)], which is also in very good agreement with the theory results [Fig.~\ref{fig:ellipDepExp}(e)].
Furthermore, both the experimentally and theoretically obtained ellipticity of the emitted $3f$-radiation $\varepsilon_{3f}$ tends to follows the driving-pulse ellipticity, as summarized in Fig.~\ref{fig:ellipDepExp}(d)(f).
These results confirm the sensitive control of the THG through ellipticity tuning, and support the interpretation of the ellipticity dependence by field-driven nonlinear intraband kinetic processes of the Dirac fermions.

To understand these ellipticity-dependent features, we scrutinise the time-dependent evolution of the electron distribution function. For an experimentally relevant relaxation time $\tau=10$~fs \cite{Kovalev2020}, snapshots
of the distribution function projected onto the ($p_\perp$,$p_\parallel$) momentum plane are presented in Fig.~\ref{fig:Distribution}(a)-(c) for representative ellipticities. 
The selected delay-times correspond to the THz fields marked by dashed lines in Fig.~\ref{fig:theoryHHG}(a)(b). The obtained time derivative of the current density is shown in Fig.~\ref{fig:Distribution}(d)(e), which is proportional to the emitted THz electric field.

For linearly polarized driving pulses [Fig.~\ref{fig:Distribution}(a)] the distribution function is strongly stretched along the field direction especially at the peak fields (see 3.87 and 4.61~ps), however the emitted electric field $\propto dj_\parallel/dt$ is nearly zero [Fig.~\ref{fig:Distribution}(d)].
In contrast, strong emission occurs when the driving THz field switches sign (e.g. at 3.52, 4.25, and 4.95~ps).
Since the time-dependent curve $\frac{dj_\parallel(t)}{dt}$ exhibits sharp peaks at these points, its overall profile deviates strongly from a $\sin$- or $\cos$-function.
This is what leads to very efficient generation of high-order harmonics. Moreover, the emitted harmonic radiation is also linearly polarized, because the perpendicular component $dj_\perp/dt = 0$ [Fig.~\ref{fig:Distribution}(e)].  

For an elliptical driving pulse [Fig.~\ref{fig:Distribution}(b),(c)], the distribution function $f(\mathbf{p})$ can be stretched along different directions as a function of time, depending on the orientation of the driving electric field. The effects of the elliptical polarization on the harmonic generation are mainly two-fold:
(i) With increasing ellipticity the current component $j_\parallel$ reduces, because the driving field component 
$E_\parallel$ decreases and, consequently, the distribution function is less stretched along this direction.
(ii) At the same time, the sharp peaks in the time-dependent curve $\frac{dj_\parallel(t)}{dt}$ become more rounded [Fig.~\ref{fig:Distribution}(d)], resulting in reduced high-harmonic generation. These two effects collaboratively cause the monotonic decrease of the harmonic yields with increasing ellipticity, as presented in Fig.~\ref{fig:theoryHHG}(c)(e) and Fig.~\ref{fig:ellipDepExp}(c)(e).

In contrast, the perpendicular HHG component experiences two competing effects. Whereas with increasing ellipticity the current density $j_\perp$ increases in favor of harmonic yielding along the same direction, its time derivative $\frac{dj_\perp(t)}{dt}$ evolves towards a $\sin$- or $\cos$-like function [Fig.~\ref{fig:Distribution}(e)], suppressing the generation of high harmonics. The maxima exhibited in the ellipticity dependent $I_\perp$ curves [Fig.~\ref{fig:theoryHHG}(d)(f) and Fig.~\ref{fig:ellipDepExp}(c)(e)] can be understood as a consequence of this competition.

In comparison with the ellipticity-dependent THG curves of the same relaxation time,
the FHG intensity $I^{(5f)}_\parallel$ decreases more rapidly with increasing ellipticity [cf. Fig.~\ref{fig:theoryHHG}(c) and Fig.~\ref{fig:theoryHHG}(e)].
Moreover, the maximum of $I^{(5f)}_\perp$ appears 
at a smaller ellipticity than for THG $I^{(3f)}_\perp$ [cf. Fig.~\ref{fig:theoryHHG}(d) and Fig.~\ref{fig:theoryHHG}(f)].
These differences show that the higher-order nonlinear effects are more sensitive to the change of the driving pulses.

In the limit case of a circular driving pulse, the distribution function is distorted by the same amount, but only the orientation rotates following the driving electric field. The corresponding current is essentially a $\sin$- or $\cos$-function of time, i.e. without high-harmonic generation.

Qualitatively, the ellipticity dependence of HHG can be obtained also analytically but for a very simplified setting (i.e. monochromatic driving pulse and perturbative regime, see Supplemental Material).
Nonetheless, these theoretical results clearly indicate that the THz high-harmonic generation in a Dirac semimetal due to intraband processes is not only very efficient, but also sensitive to the ellipticity of the driving pulses.

In conclusion, we obtain a very efficient control of terahertz third-harmonic yield and polarization state in thin films of the three-dimensional Dirac semimetal Cd$_3$As$_2$ via tuning ellipticity of the fundamental frequency.
The sensitive dependence of the high-harmonic yields on the ellipticity can be understood in terms of terahertz field driven intraband kinetics of massless Dirac fermions, which are characterized by a linear dispersion relation. 
Our study paves the way for realizing novel nonlinear photonic devices in few terahertz frequency range based on Dirac or Weyl semimetals, where terahertz high-harmonic generation and its ellipticity tunability could be exploited for signal processing and optical communications.

\begin{acknowledgments}
We thank Jens Koch for technical support. The work in Cologne was partially supported by the DFG (German Research Foundation) via Project No. 277146847 — Collaborative Research Center 1238: Control and Dynamics of Quantum Materials (Subproject No. B05).
The work in Dresden was supported in part by the DFG  through ct.qmat (EXC 2147, project-id 390858490).
F.X acknowledges the support from the National Natural Science Foundation of China (52150103, 11934005, and 11874116), National Key Research and Development Program of China (Grant No. 2017YFA0303302 and 2018YFA0305601), the Science and Technology Commission of Shanghai (Grant No. 19511120500), the Shanghai Municipal Science and Technology Major Project (Grant No. 2019SHZDZX01), the Program of Shanghai Academic/Technology Research Leader (Grant No. 20XD1400200).
P.S. acknowledges the support of the Narodowe Centrum Nauki (NCN) Sonata Bis grant 2019/34/E/ST3/00405 and the Nederlandse Organisatie voor Wetenschappelijk Onderzoek (NWO) Klein grant via NWA route 2.
Z.W. acknowledges support by the European Research Council (ERC) under the Horizon 2020 research and innovation programme, grant agreement No. 950560 (DynaQuanta).

S.G. and R.M.A.D. contributed equally to this work.

\end{acknowledgments}

\bibliography{paper_references_arxiv}

\section*{Supplemental Material}

\renewcommand{\thefigure}{S\arabic{figure}}
\setcounter{figure}{0}

\begin{figure}[h]
\includegraphics[width=5cm]{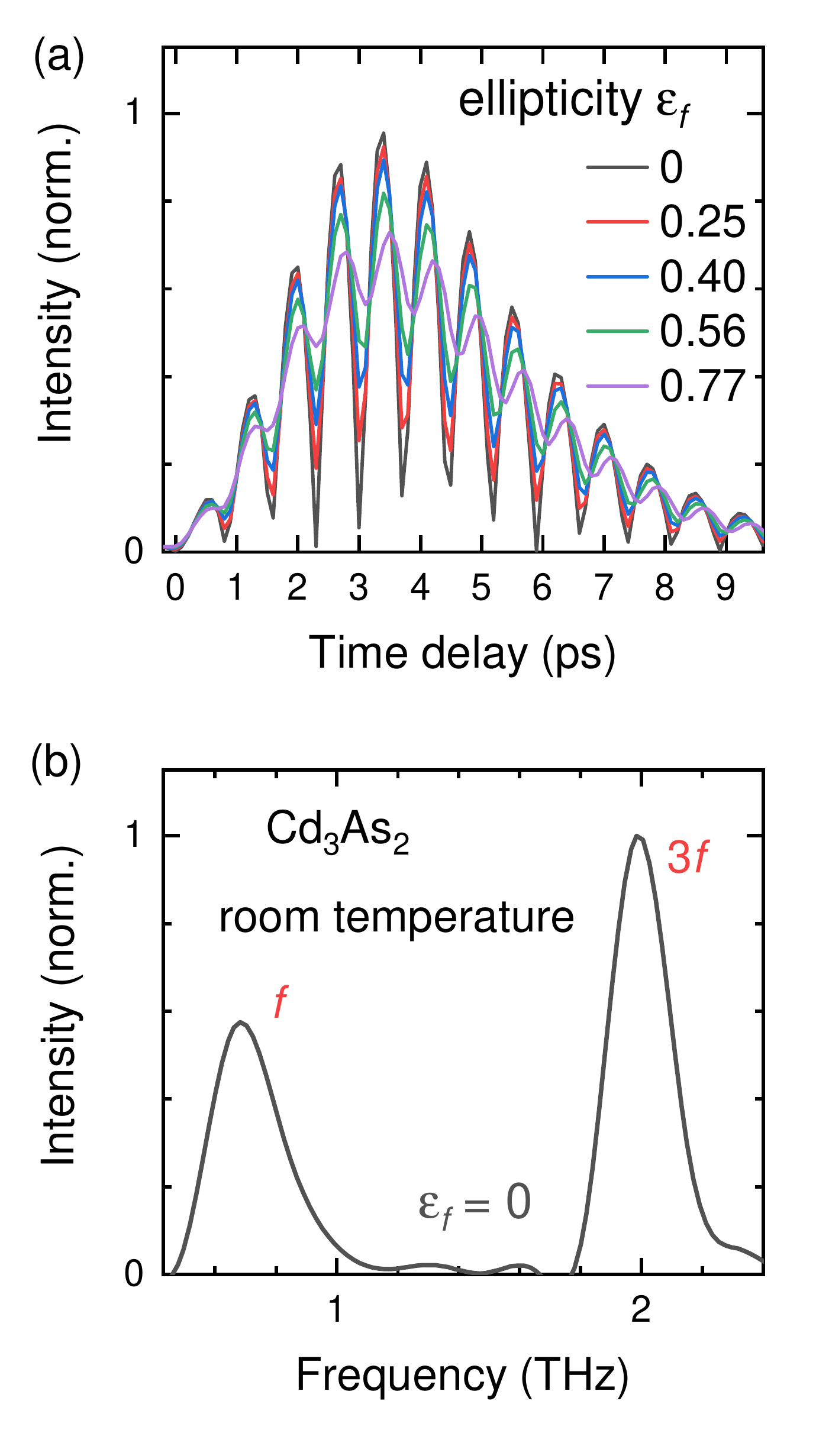}
\caption{\label{fig:Ext1}
(a) Intensity of the experimentally realized driving pulses with various ellipticities.
(b) Spectrum of emitted terahertz radiation from Cd$_3$As$_2$ for linearly polarized driving pulse ($\varepsilon_f=0$), measured through a $3f$-bandpass filter.
}
\end{figure}

\textbf{Analytical perturbative theory}. In the perturbative regime \cite{Dantas21}, one can perform an analytical analysis at zero temperature for a sinusoidal pulse $\mathbf{E} (t)= \tfrac{E }{\sqrt{\varepsilon_f^2+1}}\left[ \varepsilon_f \cos(\omega t) \,\hat{\mathbf{e}}_\perp + \sin(\omega t) \,\hat{\mathbf{e}}_\parallel \right]$ with $\omega=2\pi f$.
In the limit of no collisions, i.e. $\tau \rightarrow \infty$, the current is simply given by 
$\mathbf{j}(t) = -2e \braket{\mathbf{v} (t,0)}$ and the intensity of the third-harmonic generation is given by
%------------------------------------------------------
\begin{equation}
\{I^{P}_{3 \omega ,\perp},I^{P}_{3 \omega ,\parallel}\} \propto
\left[\frac{3 \, \kappa \, e}{80 \pi^2 f^2}  \left( \frac{v_F e \, E}{\mu}\right)^3 \right]^2  \frac{(\varepsilon^2_f-1)^2}{(\varepsilon_f^2+1)^3} \{ \varepsilon_f^2,1\},
\end{equation}
%------------------------------------------------------
where $\kappa= \tfrac{\mu^3}{6 \pi^2 \hbar^3 v^2_F}$.
This means that for a monochromatic driving pulse, the parallel component $I^{P}_{\parallel,3 f}$ of THG drops monotonically with increasing ellipticity, while the perpendicular component $I^{P}_{3f,\perp}$ reaches a maximum for
$\varepsilon_f^{\text{max}}=\frac{1}{\sqrt{5}}\approx 0.45$.
If collisions with a finite relaxation time $\tau$ (with $\tau \ll 1/f$) are taken into account, the THG intensity becomes
%------------------------------------------------------
\begin{equation}
\{I^{P,\tau}_{3 \omega ,\perp},I^{P,\tau}_{3 \omega ,\parallel}\} \propto
 \frac{36 }{\left(36+\frac{49}{\tau^2 \omega^2}+\frac{14}{\tau^4 \omega^4} +\frac{1}{\tau^6 \omega^6}\right)} \{I^{P}_{3 \omega ,\perp},I^{P}_{3 \omega ,\parallel}\}.
\end{equation}
%------------------------------------------------------
On the one hand, the third-harmonic yield reduces with increasing scattering rate $1/\tau$. On the other hand, for a fixed $1/\tau$, the maximum of $I^{P}_{3f,\perp}$ occurs at the same ellipticity
$\varepsilon_f^{\text{max}}=\frac{1}{\sqrt{5}}$.
We should note that these analytical results are valid for an homogenous electric field in the perturbative regime, which for Cd$_3$As$_2$ means an THz electric field of $E\lesssim 5$~kV/cm \cite{Dantas21,pnas.2200367119}. 
With a typical peak THz electric field of 100~kV/cm,
our experiment deals with the non-perturbative regime \cite{Dantas21}, for which the ellipticity dependence is also a function of the THz electric field, the driving-pulse waveform, and the relaxation time, therefore we have to solve the problem numerically as presented in the main text.

\end{document}